\documentclass[a4paper,11pt]{article}
\usepackage{jcappub} % for details on the use of the package, please see the JINST-author-manual
\usepackage{lineno}

\title{\boldmath Dark I-Love-Q}

\author[a,b]{Jing-Yi Wu,}
\author[b]{Wei Li,}
\author[b]{Xin-Han Huang}
\author[b,c,d]{and Kilar Zhang}
\affiliation[a]{School of Astronomy and Space Science, University of 
Chinese Academy of Sciences (UCAS), Beijing 100049, China}
\affiliation[b]{Department of Physics and Institute for Quantum Science and Technology, Shanghai University, 99 Shangda Road, Shanghai 200444, China}
\affiliation[c]{Shanghai Key Lab for Astrophysics, Shanghai 200234, China}
\affiliation[d]{Shanghai Key Laboratory of High Temperature Superconductors, Shanghai 200444, China}

\emailAdd{kilar@shu.edu.cn}

\abstract{For neutron stars, there exist universal relations insensitive to the equation of states,  the so called I-Love-Q relations, which show the connections among the moment of inertia, tidal Love number and quadrupole moment. In this paper, we show that these relations also apply to dark stars, bosonic or fermionic.  The relations can be extended to higher ranges of the variables,  clarifying the deviations for dark stars in the literature, as those curves all approximate the ones generated by a polytropic equation of state,  when taking the low density (pressure) limit. Besides, we find that for equation of states with scaling symmetries, the I-Love-Q curves do not change when adjusting the scaling parameters.}

\begin{document}
\maketitle
\flushbottom

\section{Introduction}
\label{sec:intro}

Since the first observation of gravitational waves (GW) from two inspiral black holes (BH) \cite{ref1},  we enter the era of multi-messenger astronomy.  
Besides BH,  GW also serve as excellent observatories for other compact stars like neutron stars (NS) \cite{ref2, ref3},  and they can read off not only the binary masses but also the tidal deformability,  which could further infer the radii.  While for traditional observation method,  the NS radius has to be calculated by X-ray burst and thermodynamics,  with some high error.  The data from GW and the NICER mission \cite{Miller:2021qha} will enhance the accuracy of our estimation about the radius.  However,  to perform this task,  the exact equation of state (EoS) of NS is needed,  which is now still an open question. Or conversely speaking, the accurate radius measurements are to narrow down the window for the true EoS. Most of the mainstream EoS models like  SLy4 \cite{ref4}, APR4 \cite{ref5} etc. have lots of parameters. Though there are some other theoretically derived ones with only one parameter like the MIT bag model \cite{ref6} or from holographic QCD \cite{Hoyos:2016zke, ref7, Li:2024ayw, Kovensky:2021kzl}, they are less realistic. Here by realistic it means that the EoS should describe a star with mass, radius and tidal deformability comparable to the observation data range.
%including most of the well-known ones like SLy4 etc. mentioned before. 
With different EoS,  the deduced NS mass, radius and tidal Love number (TLN) vary and show strong dependence.   However, in 2013 some universal relations insensible to EoS was found \cite{ref8, ref9}, which showed the relations among the (reduced dimensionless) momentum of inertia I,  TLN and quadrupole moment Q. These are called the I-Love-Q relations. They show that, any two of the three variables form a function which is almost independent of the EoS applied, as long as it is realistic. Thus I-Love-Q relations can serve as a quick investigation of the observation data, to infer one of the trio from another, or to test the validity of general relativity \cite{ref10,ref11}.

In the astronomical observation for NS and BH, there are so-called gap events distinguished by star masses,  with the lower gap between $2.5$ and $5$ $M_\odot$ (solar mass),  and the higher gap between $80$ and $150$ $M_\odot$.  For the lower gap,  2.5 $M_\odot$ is too heavy for an NS,  while $5$ $M_\odot$ is in general too light for a stellar BH, unless it is a primordial BH, or a remnant of other NS merger event. For the higher gap, this range is not allowed for stellar BH, so the most possibility is primordial BH,  or a remnant of other BH merger.  Nonetheless, there might be another candidate for gap events, the dark stars (DS).  Dark matter (DM) is generally believed to be dispersed, but considering the fact that it occupies $27\%$ of the universe energy,  much larger than the $5\%$ contribution from normal matter, we cannot exclude the possibility that it could form dense stars \cite{Kouvaris:2015rea, Eby:2015hsq}. The star formation might be through Bondi accretion \cite{ref12, ref13, ref14, ref15} or other process, which we choose not to touch here and leave for the future work. 

If DS do exist,  then what are their EoS? Do they satisfy similar I-Love-Q relations or not? There are indeed many DM models, fermionic or bosonic,  and the corresponding EoS can be extracted in principle.  However, usually this is not an easy process \cite{ref16},  where numerical approximations or shooting method etc. have to be applied.  For boson stars (BS),  in \cite{ref17, ref18} a new approach using isotropic limit is applied, then the EoS for different models can be systematically obtained. Although it is called a "limit",  this limit is also a necessary condition to use the Tolman-Oppenheimer-Volkoff (TOV) equations anyway, so it is actually an accurate analytic approach. In this paper,  we calculate the  I-Love-Q relation for DS, using the BS EoS obtained in \cite{ref18},  and fermion stars (FS) EoS in \cite{ref19}. Those for NS are also reproduced as references. We find that DS also enjoy similar  I-Love-Q relations,  and in fact that their curves coincide with the NS case. This is also interesting in the sense that  I-Love-Q relations are more universal than expected,  which apply to a large range of compact star (there are works on white dwarfs \cite{boshkayev2016love, taylor2020love, boshkayev2018non, roy2021universal}).  One interpretation for why this happens is, the I-Love-Q variables are mainly affected by the outer layer of the stars, which means the energy density $\rho$ and pressure $p$ are small, while in this limit all those EoS become alike.
It is worth mentioning that the universal relations are violated a lot when there exists a strongly one order phase transition at low densities \cite{ref20,ref21}.

Although the I-Love-Q relations have been verified for many continuous EoS \cite{Maselli:2013mva, Haskell:2013vha, Silva:2017uov, Adam:2020aza, Li:2023owg, Atta:2024ckt, Kumar:2023ojk}, there is no insurance that any continuous EoS will give the same results. Recently there seems to be an opinion that any smooth EoS (without strong discontinuity) satisfies the universal relation. But as is already pointed out in the original paper \cite{ref9}, that even for a ploytropic EoS with gamma=0.75 (or equivalently n=3 \footnote{There are two other conventions often encountered: $p=\kappa \rho^{\gamma}$ or $p=\kappa \rho^{1+1/n}$. }), there is an obvious violation of I-Q relation by 20$\%$ while 10$\%$ for I-Love and Q-Love relations. Also, deviations will occur when applying some extreme models, and this conclusion is verified in \cite{Yagi:2014qua, Pani:2015tga}. Therefore, it is worthy and necessary to check whether a new EoS (continuous or not) will obey the I-Love-Q relations.

The organization of the paper is as follows. Section~\ref{sec:eos} introduces the EoS used; section~\ref{sec:scsy} discusses scaling symmetries and gap events; section~\ref{sec:iloveq} shows the extended I-Love-Q relations for both DS and NS. We conclude in section~\ref{sec:conclu}.

\begin{figure}
    \includegraphics[width=1\textwidth]{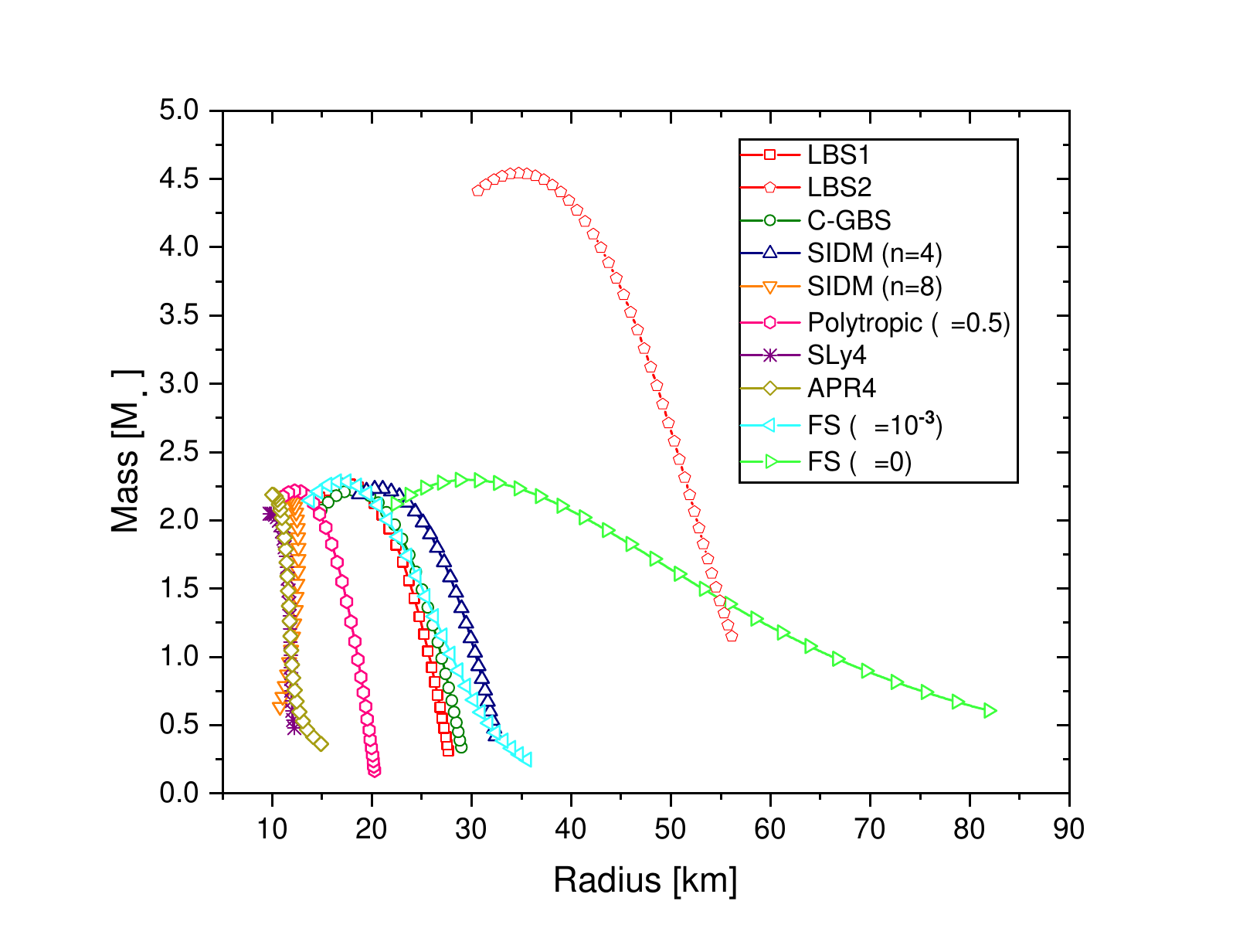}
    \caption{Mass-radius relations of 5 BS models, 2 FS models and 3 NS models. The parameters for all adjustable models are set as follows: $K=5.3 \times 10^{-4}$ in eq.\eqref{lbs} for Liouville field stars (LBS1) while $K=1.325 \times 10^{-4}$ for the curve twice in scale (LBS2); $K=1.6 \times 10^{-4}$ in eq.\eqref{eq3} for C-GBS model; $K_4=5.5 \times 10^{-2}$ 
    in eq.\eqref{eq2} for $\phi_4$ BS (SIDM $(n=4)$) and $K_8=7.6 \times 10^{-3}$ for $\phi_8$ BS (SIDM $(n=8)$) ; $\kappa=0.09$ for Polytropic $(\gamma=0.5)$ NS; we set $m_\phi=m_X$ while $m_X^4 = 0.386$ in eq.\eqref{fs} for FS $(\alpha = 10^{-3})$ model and $m_X^4=0.028$ for FS $(\alpha=0)$ model. The masses of the Polytropic $(\gamma=0.5)$, FS $(\alpha=0)$ and the 5 BS models can be easily lifted to gap events range by applying scaling symmetries \cite{ref18, ref24}.}
    \label{fig1}
\end{figure}

\begin{figure}
    \includegraphics[width=1\textwidth]{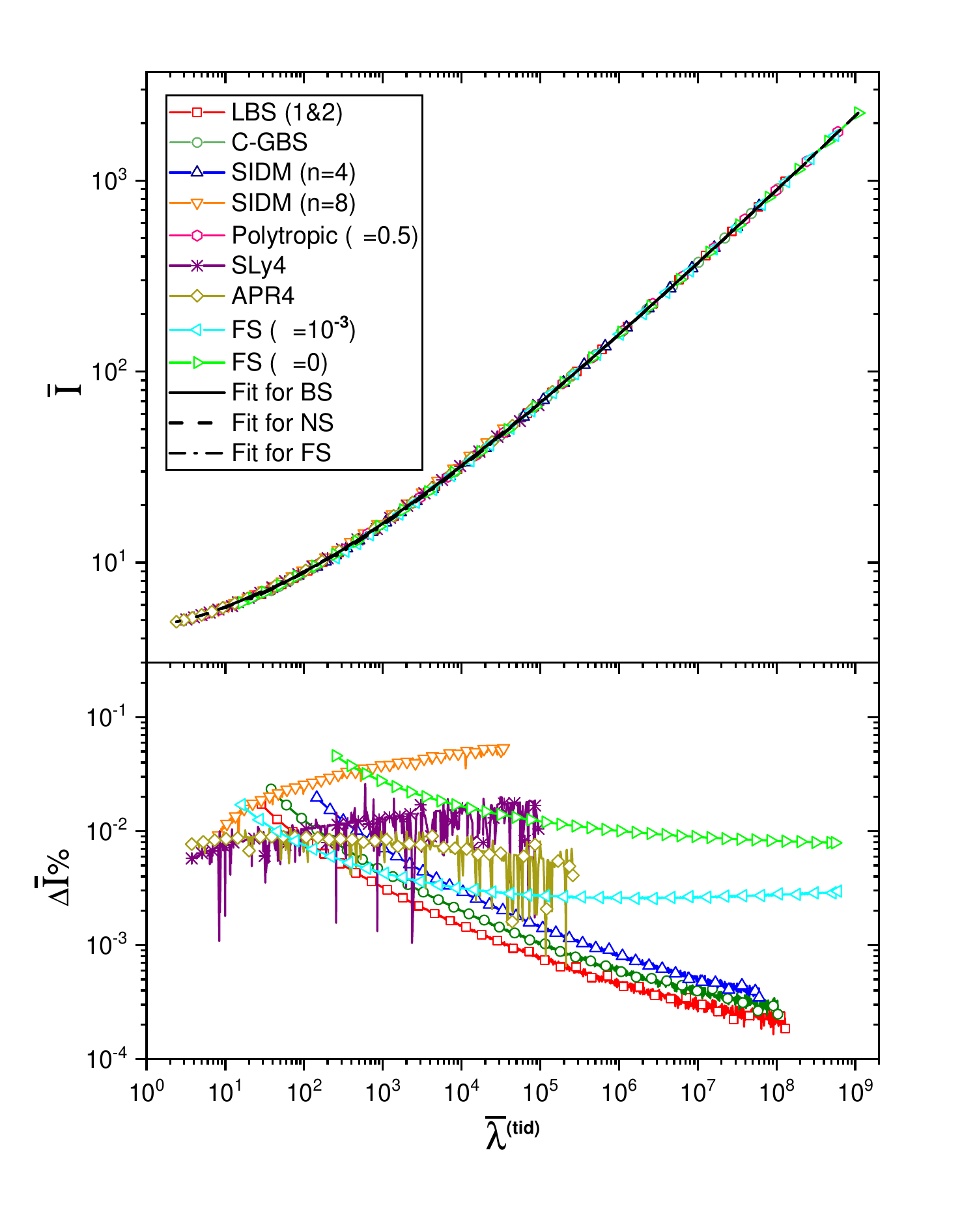}
    \caption{Top: I-Love relations of 5 BS models, 2 FS models and 3 NS models. Notice that the two LBS curves coincide. The black solid line refers to the fitting for BS, the black dotted line for FS,  and the black dashed line represents the fitting for NS. Bottom: The relative fractional errors for I-Love relations whose referenced line is polytropic NS model with index $\gamma=0.5$.}
    \label{fig2}
\end{figure}

\begin{figure}
    \includegraphics[width=1\textwidth]{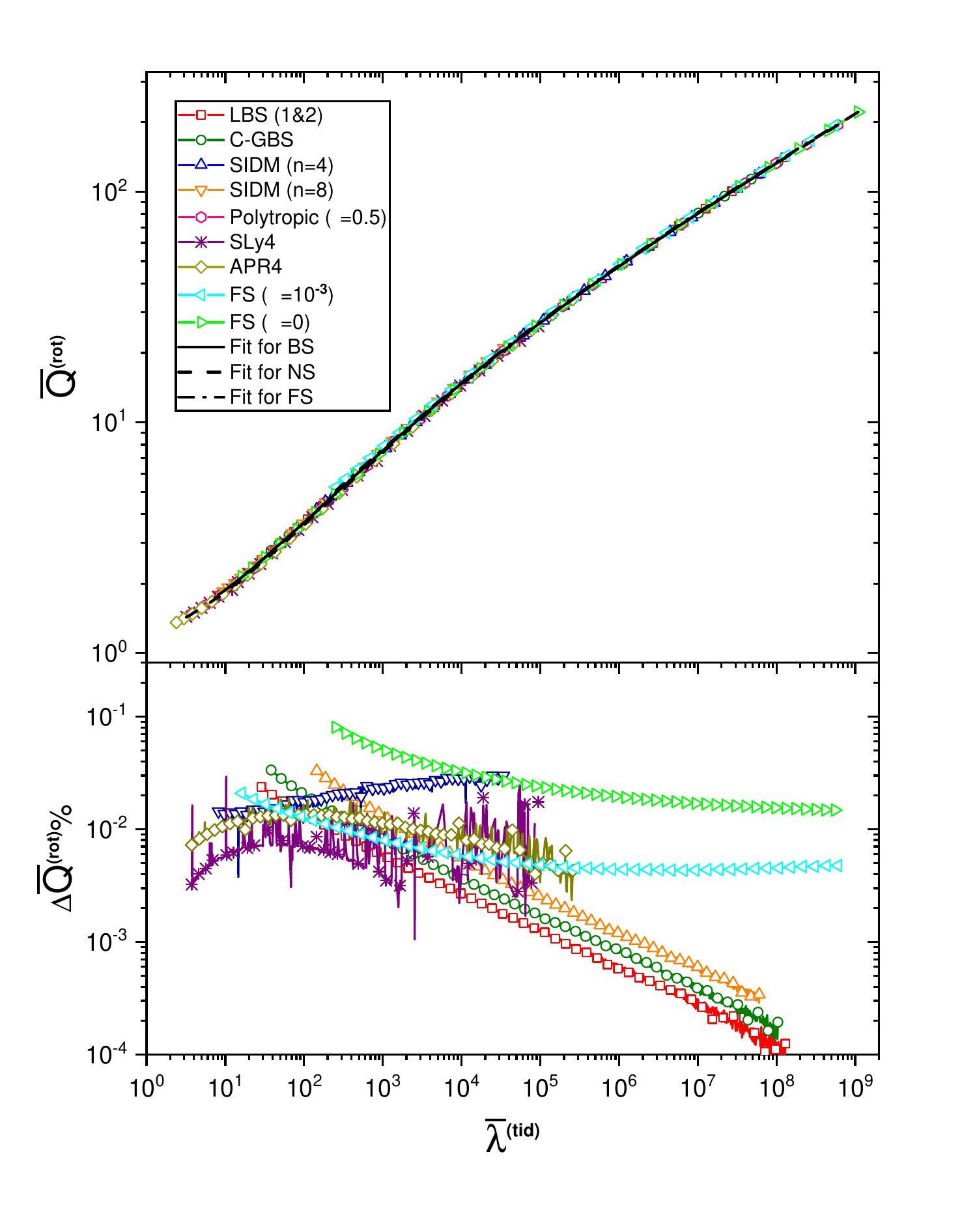}
    \caption{Top: Q-Love relations of 5 BS models, 2 FS models and 3 NS models. The two LBS curves coincide. The meanings of different lines are the same as those in figure~\ref{fig2}. Bottom: The relative fractional errors for Q-Love relations whose referenced line is polytropic NS model with index $\gamma=0.5$.}
    \label{fig3}
\end{figure}

\begin{table*}
    \renewcommand\arraystretch{1.5}
    \centering
    \begin{tabular}{|cc|ccccc|}
        \hline
        $ y_i$ & $ x_i$ & $a_i$ & $b_i$ & $c_i$ & $d_i$ & $e_i$ \\
        \hline
        $\bar{I}$ & $\bar{\lambda}^{(tid)}$ & 1.57 & 3.18$\times 10^{-2}$ & 2.70$\times 10^{-2}$ & -1.02 $\times 10^{-3}$ & 1.61$\times 10^{-5}$ \\
        $\bar{I}$ & $\bar{Q}^{(rot)}$ & 1.46 & 0.456 & 4.42 $\times 10^{-2}$ & 3.15 $\times 10^{-2}$ & -2.86$\times 10^{-3}$\\
        $\bar{Q}^{(rot)}$ & $\bar{\lambda}^{(tid)}$  & -7.73$\times 10^{-3}$ & 0.243 & 1.44$\times 10^{-2}$ & -1.26$\times 10^{-3}$ & 2.97$\times 10^{-5}$
    \\
        \hline
    \end{tabular}
    \caption{Numerical fitting for the I-Love, Q-Love and I-Q relations of the 5 BS models.}
    \label{tab1}
\end{table*}
\begin{table*}
    \renewcommand\arraystretch{1.5}
    \centering
    \begin{tabular}{|cc|ccccc|}
        \hline
        $ y_i$ & $ x_i$ & $a_i$ & $b_i$ & $c_i$ & $d_i$ & $e_i$ \\
        \hline
        $\bar{I}$ & $\bar{\lambda}^{(tid)}$ & 1.53 & 4.10$\times 10^{-2}$ & 2.50$\times 10^{-2}$ & -8.51 $\times 10^{-4}$ & 1.17 $\times 10^{-5}$ \\
        $\bar{I}$ & $\bar{Q}^{(rot)}$ & 1.44 & 0.483 & 1.34$\times 10^{-2}$ & 4.12$\times 10^{-2}$ & -3.77$\times 10^{-3}$ \\
        $\bar{Q}^{(rot)}$ & $\bar{\lambda}^{(tid)}$ & -1.82$\times 10^{-1}$ & 0.325 & 1.86$\times 10^{-3}$ & -4.60 $\times 10^{-4}$ & 1.15$\times 10^{-5}$ \\
        \hline
    \end{tabular}
    \caption{Numerical fitting for the I-Love, Q-Love and I-Q relations of the 2 FS models.}
    \label{tab2}
\end{table*}
\begin{table*}
    \renewcommand\arraystretch{1.5}
    \centering
    \begin{tabular}{|cc|ccccc|}
        \hline
        $ y_i$ & $ x_i$ & $a_i$ & $b_i$ & $c_i$ & $d_i$ & $e_i$ \\
        \hline
        $\bar{I}$ & $\bar{\lambda}^{(tid)}$ & 1.52 & 5.45$\times 10^{-2}$ & 2.31$\times 10^{-2}$ & -7.52 $\times 10^{-4}$ & 9.89$\times 10^{-6}$ \\
        $\bar{I}$ & $\bar{Q}^{(rot)}$ & 1.41 & 0.574 & -2.18$\times 10^{-2}$ & 4.58$\times 10^{-2}$ & -3.96$\times 10^{-3}$ \\
        $\bar{Q}^{(rot)}$ & $\bar{\lambda}^{(tid)}$ & 8.20$\times 10^{-2}$ & 0.193 & 2.14$\times 10^{-2}$ & -1.61 $\times 10^{-3}$ & 3.53$\times 10^{-5}$ \\
        \hline
    \end{tabular}
    \caption{Numerical fitting for the I-Love, Q-Love and I-Q relations of the 3 NS models.}
    \label{tab3}
\end{table*}

\section{Equation of State}
\label{sec:eos}

To accurately derive the effects of rotation and tidal forces on relativistic stars, establishing the EoS is a crucial first step. 
Observational evidence has led to the proposal of various dark matter models, which can be broadly classified into two categories based on the properties of their constituent particles: FS and BS.  In the formation of compact stars, 
FS are stabilized against gravitational collapse by the degenerate pressure exerted by fermions whereas BS rely on the self-interaction to maintain their structure. Yet for FS to have a considerable size, self-interaction should also be introduced.

For NS,  we employ two widely used EoS models,  SLy4 and APR4, together with a simple polytropic model,   $\rho=\kappa p^{\gamma}$, with $\gamma=\frac{1}{2}$ .

In the case of DS,  we examine four different self-interacting dark matter (SIDM) models, comprising three bosonic \cite{ref18} and one fermionic \cite{ref19}.
The first one describes BS with a scalar potential given by $V_n(\phi)=\frac{m^2}{2}|\phi|^2+\frac{\lambda_n}{n \Phi^{n-4}_0}|\phi|^n$,
where $\lambda_n$ is a dimensionless coupling constant, and $\Phi_0$ is a constant with the same dimension as the potential $\phi$ \cite{ref23}. Upon applying the isotropic limit method, 
the corresponding EoS is derived \cite{ref18}
\begin{equation}
    \rho=\frac{n+2}{n-2} p + K_n p^{\frac{2}{n}},\label{eq2}
\end{equation}
where $K_n=\left( \frac{2n}{n-2} \right)^{\frac{2}{n}}\left( \frac{m^2 M_{pl}^2}{4\pi \Lambda_n} \right)^{1-\frac{2}{n}}$. $\Lambda_n$ is defined as
$\Lambda_n=\left(\lambda_n \frac{\Phi_{0}^{2}}{m^2}\right)^{\frac{2}{n-2}} \frac{M_{pl}^2}{\Phi_{0}^{2}}$, and $M_{pl}$ means
the Planck mass. It requires $\Lambda_n \gg 1$ when achieving the isotropic limit.  It is evident that as $p \to 0$, this EoS approximates a polytropic model with $\gamma=\frac{2}{n}$.

The second model is Liouville field associating with the potential $V(\phi)=\frac{m^2}{2\varepsilon^2}[e^{\varepsilon^2 |\phi|^2}-1]$. The corresponding EoS for this model is introduced in \cite{ref18}
\begin{equation}
    \begin{split}
        \rho=K(\eta^2 e^{\eta^2} + e^{\eta^2} -1), \\
        p=K(\eta^2 e^{\eta^2} - e^{\eta^2} +1), \label{lbs}
    \end{split}
\end{equation}
where $\eta$ is the variable and $K$ is defined as $K=\frac{m^2 M_{pl}^2}{4\pi \Lambda}$ with $\Lambda = \varepsilon M_{pl}$.
In the isotropic limit, analogous to $\Lambda_n$, it is necessary that $\Lambda \gg 1$. As $p \to 0$, this EoS approximates $\rho  \propto p^\frac{1}{2}$.

The last bosonic model is cosh-Gordon field with $V(\phi)=\frac{m^2}{\varepsilon^2} \left[\cosh(\varepsilon \sqrt{|\phi|^2})-1 \right]$ whose
EoS is \cite{ref18}
\begin{equation}
    \begin{split}
        \rho=K \left(\frac{1}{2}\eta \sinh\eta + \cosh \eta -1 \right), \\
        p=K \left(\frac{1}{2}\eta \sinh \eta -\cosh \eta +1 \right),  \label{eq3}
    \end{split}
\end{equation}
where $K$ retains the same definition as in eq.\eqref{eq3}, and the isotropic limit condition is invariant for $\Lambda$. When $p \to 0$, this EoS also approaches $\rho  \propto p^\frac{1}{2}$.

For fermionic models,  we apply the EoS from \cite{ref19}, which considers the Yukawa potential $V={\alpha \over r} \exp(-m_\phi r)$ where $\alpha$ represents the dark fine structure constant and $m_\phi$ denotes the mediator mass:
\begin{equation}
    \begin{split}
    \rho=&\frac{m_X^4}{8 \pi^2}\left[ x\sqrt{1+x^2}(2x^2+1) -\ln(x+\sqrt{1+x^2}) \right]+\frac{2 \alpha m_X^6}{9 \pi^3 m_{\phi}^2}x^6\;, \\
    p=&\frac{m_X^4}{8 \pi^2}\left[ x\sqrt{1+x^2}(2x^2/3-1) +\ln(x+\sqrt{1+x^2}) \right]+\frac{2 \alpha m_X^6}{9 \pi^3 m_{\phi}^2}x^6. \label{fs}
    \end{split}
\end{equation}
Here $m_X$ is the fermion mass, and $x$ denotes the dimensionless Fermi momentum. As $p \to 0$, this EoS approaches $\rho  \propto p^\frac{3}{5}$ .

\section{Mass-Radius relations and Scaling Symmetries}
\label{sec:scsy} 

In Figure~\ref{fig1} we show the mass-radius relations of five BS models and two FS models.
Additionally, we compare them with NS using a polytropic EoS with $\gamma=\frac{1}{2}$ as well as two
realistic models, SLy4 and APR4. Each relation is plotted by varying the central pressure of the stars. 
Considering DS could be neutron star mimickers through existing observing manners, the upper mass limit for all but one of these models are adjusted to around 2.2$M_\odot$, the observational limit for NS. 

We are also interested in the mass range of gap events, so we also add one EoS with the maximum mass of about 4.5$M_\odot$, within the lower mass gap. 
Notice that for the polytropic model, FS $(\alpha=0)$ model and the 5 BS models, there are scaling symmetries \cite{ref18, ref24}, which makes their mass-radius curves easily be magnified in direct proportion.

In details, the scaling symmetries guarantee that, as long as $\rho$ and $p$ can be re-scaled simultaneously by $k$ for a given EoS (this condition holds for most analytic EoS), the corresponding star mass $M$ and radius $R$ will be simultaneously scaled by ${1\over {\sqrt k}}$. Then the "compactness" $C=M/R$ remains invariant. It is proved in \cite{ref18} that the dimensionless TLN is also invariant. Here with the same reasoning, we find that all dimensionless variables will stay unchanged, since the scaling parameters will be canceled when making a variable dimensionless. That is to say, the dimensionless trios (the momentum of Inertia I,  TLN and quadrupole moment Q) are all invariant under scaling symmetries.

\begin{figure}
    \includegraphics[width=1\textwidth]{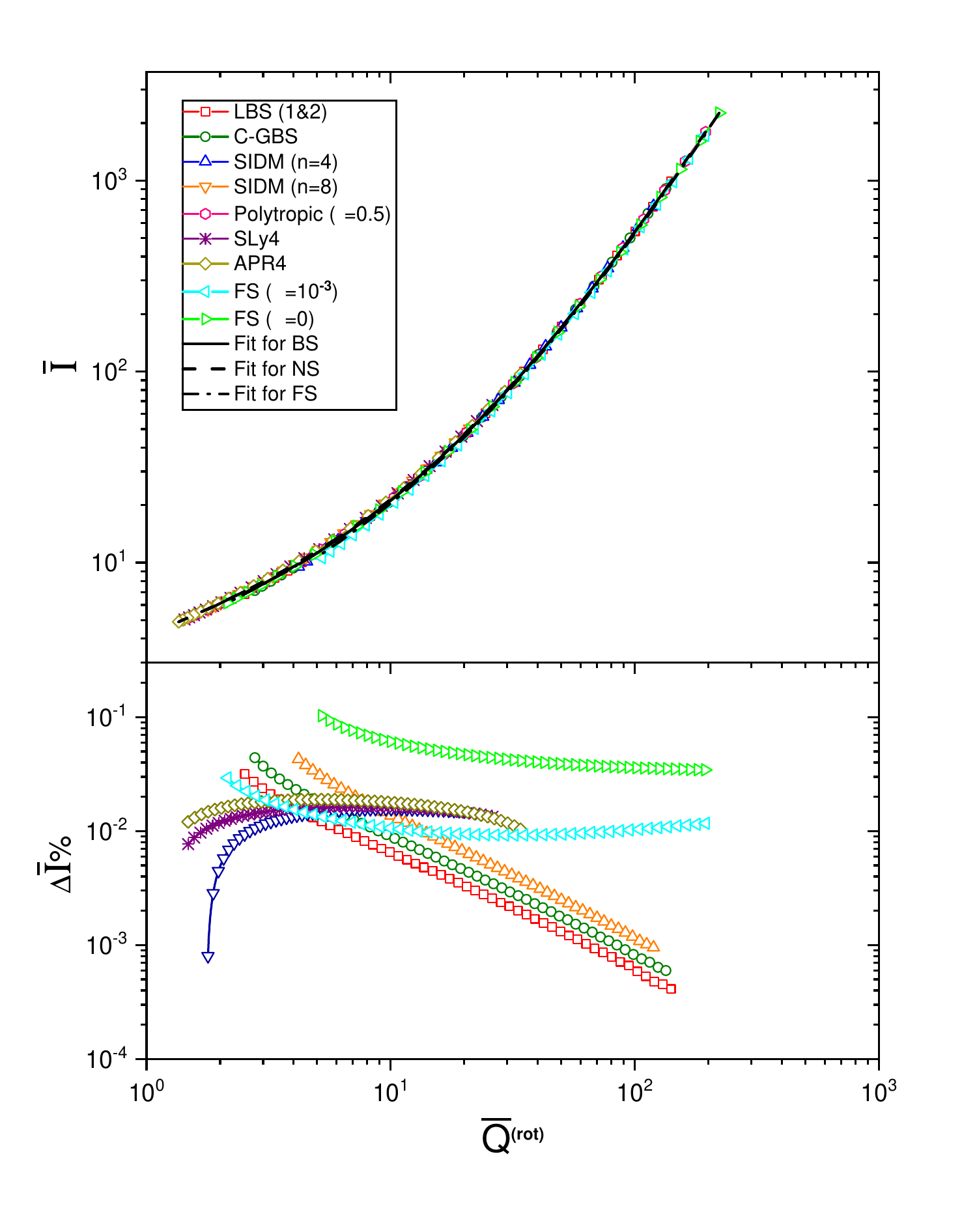}
    \caption{Top: I-Q relations of 5 BS models, 2 FS models and 3 NS models. The two LBS curves coincide. The meanings of different lines are the same as those in Fig. \ref{fig2}. Bottom: The relative fractional errors for Q-Love relations whose referenced line is polytropic NS model with index $\gamma=0.5$.}
    \label{fig4}
\end{figure}

As a result, the scaling symmetries allow us to interpret the gap events as DS with several or tens of solar masses, without affecting the I-Love-Q relation curves. Although in Fig. \ref{fig1} the Mass-Radius curves are given for mostly the neutron star range, we should bear in mind that those curves can scan all the mass ranges needed, including the lower and higher mass gaps. In order to show the proportion, we plot two curves for Liouville field stars, with one having twice mass-radius values of the other, labeled as "LBS2" and "LBS1", respectively. The invariance of I-Love-Q trios induced by scaling symmetries is exhibited in Figure~\ref{fig2} to \ref{fig4} in the next section.

%\newpage
\section{I-Love-Q relations}
\label{sec:iloveq}

With the techniques introduced in \cite{ref9},  we can calculate the moments of inertia $\bar{I}$,  TLN $\bar{\lambda}^{(tid)}$ and quadrupole moments $\bar{Q}^{(rot)}$ for DS,  along with those for NS serving as references. The "bar"s above the three variables indicate that they are already made dimensionless.
The top panels of Figure~\ref{fig2}, Figure~\ref{fig3} and Figure~\ref{fig4} show the I-Love, Q-Love and I-Q relations of these models, respectively.
In addition, we make fittings for BS, FS and NS, with their I-Love-Q trios fitted on a log-log scale in a form as in \cite{ref8}:
\begin{equation}
    \ln y_i=a_i + b_i \ln x_i+ c_i (\ln x_i)^2 + d_i (\ln x_i)^3 +e_i (\ln x_i)^4.
\end{equation}
In Table~\ref{tab1}, \ref{tab2} and \ref{tab3}, we list $a_i$, $b_i$, $c_i$, $d_i$ and $e_i$ for BS, FS and NS models, which  respectively referred to the black solid
line, dash-dot line and dashed line in Figure~\ref{fig2} to \ref{fig4}.

We find that, the original I-Love-Q relations also apply to BS and FS, with some slight deviation when $\bar{\lambda}^{(tid)} <200$ (equivalently $\bar{I}< 10$ or $\bar{Q}< 4$).
One interpretation for this phenomenon is that, those three parameters are mainly affected by the outer core of the compact stars, where the pressure and density are low. And all those DS EoS reduce to similar polytropic ones with $\gamma \leq 0.6$ when $p \to 0$.  On the other hand,  for small TLN, the stars are more compact and smaller in size, and the pressure is higher, and different EoS have different forms, which leads to more distinctions. Moreover, we verified the trios for ``LBS1'' and ``LBS2'' separately and the results fully agree with the previous description about scaling symmetries. So we use the figure legend ``LBS 1\&2'' to represent the I-Love-Q relations of Liouville BS model.  

While the original relations in \cite{ref8} are fitted up to approximately $ \bar{\lambda}^{(tid)}=2\times 10^4$, the upper range for NS,  here the upper range is set to $10^8$ like in \cite{ref24} and \cite{ref9},  as DS might have larger TLN. 
 We must emphasize that, in \cite{ref24} it shows that the FS curves start deviating from the original I-Love-Q relations when $\bar{\lambda}^{(tid)} >10^5$.  However, this arise from the fact that, the original curve in \cite{ref8} is fitted to a smaller range,  so directly extending the semi-analytic fit in \cite{ref8} to high TLN will naturally not be accurate enough. In fact,  when we fit the NS EoS up to $ \bar{\lambda}^{(tid)}= 10^8$,  they coincide with the BS and FS curves also at large TLN range, as shown in Figure~\ref{fig2} to \ref{fig4}.

Bottom panels of these figures are the relative fractional errors calculated against the polytropic NS EoS with $\gamma = \frac{1}{2}$ as referenced curves (to be more precise, one can choose the numerical fit curves listed in Table~\ref{tab1} to \ref{tab3} as referenced curves like in \cite{ref8}, but it is more convenient to use the polytropic EoS). We see that the errors are small except for the higher compactness part of FS with $\alpha=0$ whose compactness is obviously smaller than other models. So we conclude that the trio violates slightly for some extreme structures, but it holds true for most cases in general. If just for simplicity, the whole I-Love-Q relations could be approximately reproduced by a single polytropic EoS with $\gamma = \frac{1}{2}$, even though this EoS is not a reasonable NS EoS, which on the contrary shows the power of EoS insensitivity of I-Love-Q relations.

\section{Conclusions}
\label{sec:conclu}
The moment of inertia, tidal Love number and quadrupole moment are insensitive to NS EoS, as proposed by Yagi and Yunes in \cite{ref8}.
In order to explore the properties of DS, we fit the I-Love-Q trio for five BS models, two FS models together with three NS models. 
It turns out that even for DS, the I-Love-Q relations still hold. The range of the original relations are expanded, and show high consistency within different models as the TLN grows (or equivalently the compactness falls), clarifying the deviations in \cite{ref24}. Besides, we confirm that the I-Love-Q relations are invariant if the EoS have scaling symmetries.

The universal relations are crucial for future observations. On the front of astrophysics, the obedience of the I-Love-Q relations for DS suggests that some of the observed GW events may originate from these NS mimickers. Therefore, further investigations into these systems, utilizing multi-messenger data, could provide valuable insights into the properties of DS. Additionally, the breakdown of these relations in extreme models offers an effective means of distinguishing between potential DM models if they do exist.      

For now, the I-Love-Q relation is a phenomenon rather than a theorem.
Special (discontinued EoS etc.) cases already show violations, but we conjecture that this relation holds for the out layer EoS takes a polytropic limit with $\gamma \leq 0.6$. One possible way leading to its proof is to evaluate how much the outer shell of the star matters, which can be verified by comparing different EoS for the outer layer with the same EoS for the inner core, and comparing different EoS for the inner core  with the same EoS for the outer layer. A test for the latter is offered in \cite{Chen:2025caq} and confirms this idea. Whether new observations support or violate this universal relation will be interesting. New runs of current terrestrial interferometers like LIGO, Virgo and KAGRA will release more data on the masses and TLN for NS and possible DS, while the future space-based detectors like eLISA \cite{Amaro-Seoane:2012vvq}, Taiji \cite{Ruan:2018tsw} and TianQin \cite{TianQin:2015yph} will bring information for more massive events which may be interpreted as DS.  The NS (or their mimickers) masses can be given by pulsar timing using radio telescope like the Five-hundred-meter Aperture Spherical radio Telescope (FAST) \cite{Nan:2011um}, and their radii (compactness) by X-ray telescope like NICER and the enhanced X-ray Timing and Polarimetry mission (eXTP) \cite{eXTP:2018anb}. In sum, I-Love-Q relation offers a new test field for general relativity and multi-messenger astronomy.

\acknowledgments

The authors thank Feng-Li Lin, Zhoujian Cao and Chen Zhang for very helpful discussions, and anonymous referees for their valuable comments and suggestions.  KZ (Hong Zhang) is supported by a classified fund from Shanghai city.

\end{document}